%% file: main.tex
\documentclass[letterpaper, 10 pt, conference]{IEEEtran}

\usepackage[noadjust]{cite}
\usepackage{graphicx,xcolor,epstopdf}
\usepackage{amsmath,amsfonts,amssymb,amsthm,mathtools}
\usepackage{subcaption}
\usepackage{booktabs}

\newcommand{\I}{\mathcal{I}}
\newcommand{\B}{\mathcal{B}}
\newcommand{\F}{\mathbb{F}}
\newcommand{\R}{\mathbb{R}}

\usepackage{hyperref}
\hypersetup{
	colorlinks,
	linkcolor={violet!50!black},
	citecolor={purple!50!black},
	urlcolor={blue!50!black}
}

\newcommand{\ra}[1]{\renewcommand\arraystretch{#1}}

\begin{document}

\title{\textit{Modelling, Simulation, and Control of a Flexible Space Launch Vehicle}}

\author{\IEEEauthorblockN{Muhammad Abdullah Aamer\IEEEauthorrefmark{1}, Qurat Ul Ain\IEEEauthorrefmark{1}, Ushbah Kaleem\IEEEauthorrefmark{1}, Hafiz Zeeshan Iqbal Khan\IEEEauthorrefmark{1}\IEEEauthorrefmark{2}, and Jamshed Riaz\IEEEauthorrefmark{1}\vspace{0.2cm}}

\IEEEauthorblockA{\IEEEauthorrefmark{1} Department of Aeronautics \& Astronautics, Institute of Space Technology, Islamabad, Pakistan.}

\IEEEauthorblockA{\IEEEauthorrefmark{2} Centers of Excellence in Science and Applied Technologies, Islamabad, Pakistan.}}

\maketitle

\begin{abstract}
\textit{\mdseries Modern Space Launch Vehicles (SLVs), being slender in shape and due to the use of lightweight materials, are generally flexible in nature. This structural flexibility, when coupled with sensor and actuator dynamics, can adversely affect the control of SLV, which may lead to vehicle instability and, in the worst-case scenario, to structural failure. This work focuses on modelling and simulation of rigid and flexible dynamics of an SLV and its interactions with the control system. SpaceX's Falcon 9 has been selected for this study. The flexible modes are calculated using modal analysis in Ansys. High-fidelity nonlinear simulation is developed which incorporates the flexible modes and their interactions with rigid degrees of freedom. Moreover, linearized models are developed for flexible body dynamics, over the complete trajectory until the first stage's separation. Using classical control methods, attitude controllers, that keep the SLV on its desired trajectory, are developed, and multiple filters are designed to suppress the interactions of flexible dynamics. The designed controllers along with filters are implemented in the nonlinear simulation. Furthermore, to demonstrate the robustness of designed controllers, Monte-Carlo simulations are carried out and results are presented.}
\end{abstract}

\begin{IEEEkeywords}
\textit{Space Launch Vehicles; Flexible Dynamics; Flexible Modes; Gain Stabilization; Notch Filters; Low Pass Filters; Elliptic Filters}
\end{IEEEkeywords}

\section{Introduction}
\input{Sections/Introduction}

\section{Mathematical Modelling}\label{sec:MathematicalModelling}
\input{Sections/Mathematical_Modelling}

\section{Control Design}\label{sec:ControlDesign}
\input{Sections/Control_Design}

\section{Nonlinear Simulation}\label{sec:NonlinearSimulation}
\input{Sections/Nonlinear_Simulation}

\section{Conclusion}\label{sec:Conclusion}
This paper considered the problem of attitude control of a flexible SLV. Falcon 9 was selected for this study. A nonlinear model of SLV dynamics (both rigid and flexible) was developed. Moreover, a classical PID type controller along with different filters was designed to mitigate the flexibility effects during the ascent phase. Specifically, notch and elliptic filters were designed and compared. The simulation results showed that the designed controller along with both filters was able to keep the SLV on the desired trajectory. Monte-Carlo analyses were also performed to compare the robustness properties of both filters. It was shown that the elliptic filter can tolerate up to $\pm 34\%$ variations in modal frequencies and mode shapes, while the notch filter could only handle $\pm 3\%$ variations.

In future, this work can be extended to further increase the fidelity of the mathematical model by incorporating the effects of inertial forces due to movement of gimballed nozzles and liquid propellant sloshing. Moreover, for off-nominal conditions and in the presence of uncertainties adaptive control algorithms can be designed to suppress the destabilizing effects.

\bibliography{References}

\end{document}

%% file: Sections/Introduction.tex
Curiosity of the mankind for space exploration has increased the need for Space Launch Vehicles (SLVs). The minimum weight objectives of large but slender SLVs have led them to exhibit structural flexibility. Structural flexibility depends upon vehicle fineness ratio, whose increase leads to issues related to vehicle dynamics and control. As flexibility increases, the modal frequencies of flexible modes get closer to the rigid body modes that may result in rigid-elastic coupling. In addition to displacement and acceleration due to rigid body motion, structural deflections can contribute to the net body motion. It is important to be able to model this change in behavior to avoid the deteriorating effects of its interference with the flight control system. This control structure coupling causes the vehicle to deviate from desired performance resulting in instability and in extreme scenarios, structural failure.

Flexibility affects the control loop in two ways. Firstly, it alters the output of the sensor to include the bending frequencies in the feedback loop and secondly, it changes the actuator position and actuator command angle resulting in altered control command to the vehicle \cite{Blakelock1991}. In general, this contribution of structural flexibility limits the control system bandwidth. Proper determination of dominant vibration modes of launch vehicles is integral in designing the attitude controller \cite{Edberg2020}. Problems due to flexibility can be avoided by modifying the control system by including filters, stiffening the sensor mounting structure or relocating sensors to locations where structural flexibility effects are minimum.


There are generally two approaches in literature for control design of a flexible SLV, i.e. gain stabilization and phase stabilization \cite{Nesline1985}. In phase stabilization, loop components and filters are selected such that the phase of structural feedback loop is $180^\circ$, whereas, in gain stabilization a filter that has deep notch is introduced at structural frequency, this keeps the loop gain well below unity. These filters such as notch filter, elliptic filter, etc. suppress vehicle's flexibility and fuel sloshing dynamics by providing gain attenuation and phase stabilization \cite{Orr2014}.

Notch filters requires prior knowledge of exact frequency of flexible modes. Since the vibration frequency varies throughout the trajectory as the propellant burns, it is difficult to calculate the exact value of flexible mode frequencies \cite{Choi2001}. This makes the use of notch filters less practical. To overcome this issue, other filters are suggested in literature, i.e. Elliptic filters \cite{Samar2008}, Kalman filters \cite{Halsey2005}, etc. Another solution is to use adaptive notch filters. Using sensor output signals, an adaptive notch filter estimates exactly the frequency of the actual system. The design parameters of the filter are updated continuously to match with the actual system parameters \cite{Oh2005}. For vehicles that have two modes close to each other, the adaptive notch filter can be extended to predict the frequency of these two modes. This type of adaptive algorithm is useful for flexible space launch vehicles that have low natural frequencies \cite{Choi2001}. For advanced launch vehicle configurations with unstable aerodynamics, high flexibility, liquid propellent sloshing, and inertia effects of engine (tail wags) the classic control methods along with filters are not effective in meeting robustness margins. For such vehicles adaptive control techniques are usually employed \cite{Smrithi2016,Pang2021}. 

In this paper, a nonlinear mathematical model for a flexible space launch vehicle is developed
incorporating the effects of flexibility at sensor and actuator locations. SpaceX’s Falcon 9 is selected as reference SLV because most of relevant data is available and remaining is obtained using CFD and Modal analysis. Based on the mathematical model, a high fidelity nonlinear simulation is developed. Furthermore, this model is linearized around a trajectory to obtain a set of linear models. Using classical control theory, linear controller is designed along with filters to mitigate the flexibility effects. In this work we have designed both notch and elliptic filters and compared their performance and robustness. The designed controller and filter are then implemented in nonlinear simulation, and results are presented. Moreover, a Monte-Carlo analysis is preformed to compare the robustness of both filters towards the uncertainties and variations in modal frequencies and mode shapes.

The rest of the paper is organized as follows. Section \ref{sec:MathematicalModelling} starts with aerodynamic and structural analysis results of Falcon 9, followed by nonlinear mathematical model of flexible SLV along with its linearized version. Thereafter, Section \ref{sec:ControlDesign} presents the designed controller and structural filters for gain stabilization of flexible modes. In Section \ref{sec:NonlinearSimulation} nonlinear simulation results are presented. Finally, the discussion is concluded in Section \ref{sec:Conclusion}. 

%% file: Sections/Mathematical_Modelling.tex
To minimize unnecessary complexity, mathematical modeling does not account for all of the variables present in the real system. To retain an accurate representation of the system, the trade-off is to make assumptions that incur minor errors in computations while considerably reducing complexity. In this work, earth is assumed to be flat and non-rotating and considered as an
inertial reference frame. Moreover, the fuel and oxidizer sloshing, and `tail wags dog' effects are ignored while developing the dynamical model of the flexible SLV.

Before proceeding towards modelling, we introduce some notation which will be used in this work. $\I$ is inertial frame, with origin at the launch point of vehicle. $\B_r$ is rigid-body fixed frame, centered at the c.g. of the SLV, with $X_B$-axis pointing towards nose and $Y_B$-axis pointing towards right. 
$\B_f$ is the local flexible-body fixed frame at the sensor location. $R_{\F_1}^{\F_2}$ represents the transformation matrix from $\F_1$ to $\F_2$, $R_{e}(\theta)$ represents the rotation of angle $\theta$ about unit vector $e$, also $e_1$, $e_2$, and $e_3$ represents the unit vectors $[1,0,0]^\top$, $[0,1,0]^\top$, and $[0,0,1]^\top$, respectively. $V$ represents velocity of c.g. of SLV in $\B_r$, and $\omega$ is angular velocity of $\B_r$ w.r.t. $\I$ expressed in $\B_r$.


\subsection{Aerodynamic and Inertial Data}
As discussed earlier, SpaceX's Falcon 9 is selected because of availability of its material and engine propulsive data from SpaceX website and other internet forums. Its slender body, with some parts made of composite structure, makes flexibility a more prominent characteristic in its behavior compared to metallic bodies of other SLVs. Since aerodynamic and inertial data is not readily available on internet, we performed CFD analysis of a CAD model of Falcon 9 from GrabCAD \cite{CAD2016}. A simplified 2D version of this CAD model is used for CFD analysis using Fluent. CFD is done at five values of angle of attack ($\alpha$) ($0^\circ$, $2^\circ$, $4^\circ$, $6^\circ$, $8^\circ$) and at five values of  Mach number ($0.5$, $1.5$, $4$, $7$, $10$). The data obtained is interpolated to obtain the values of $C_L$, $C_D$, and $C_m$, i.e. lift, drag, pitching moment coefficients, respectively, at each point in the trajectory. Moreover, symmetry of the SLV shape is exploited to obtain the aerodynamic data for negative values of angle of attack, and directional coefficients as a function of side-slip angle ($\beta$) and Mach number. Moreover, the rate derivatives e.g. ${C}_{m\dot{\alpha}}$ are assumed to be negligible and the dynamic derivatives ($C_{m_{q}},C_{n_{r}},C_{l_{p}}$) are obtained from \cite{Delma1980}.

\begin{figure}[b]
    \centering
    \includegraphics[width=0.9\linewidth]{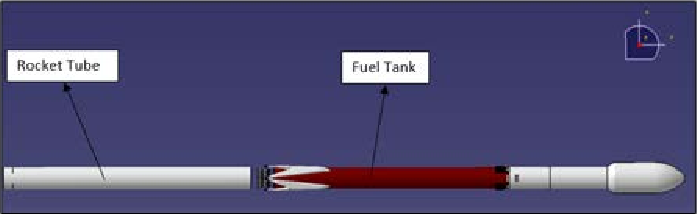}
    \caption{Exploded view of CAD model showing fuel tank}
    \label{fig:ExplodedView}
\end{figure}

\ra{1.3}
\begin{table*}
\centering
\caption{Estimated inertial data at different fuel conditions}
\label{table:InertialData}
\begin{tabular}{cccccc}\toprule
\textbf{Parameter} & \textbf{100\% fuel} & \textbf{75\% fuel} & \textbf{50\% fuel} & \textbf{25\% fuel} & \textbf{0\% fuel}\\ \cmidrule(lr){1-6}
mass (kg) & 581726.686 & 511784.213 & 441841.74 & 371899.268 & 301956.795\\
CG [from nose] (mm) & 38192.31 & 38939.239 & 38542.382 & 36340.671 & 31093.342\\
$J_{xx}$ (kgm\textsuperscript{2}) & $1.516\times 10^{6}$ & $1.408\times 10^{6}$ & $1.299\times 10^{6}$ & $1.191\times 10^{6}$ & $1.083\times 10^6 $ \\
$J_{yy}$, $J_{zz}$ (kgm\textsuperscript{2}) & $2.545 \times 10^{8}$ & $2.516\times 10^{8}$ & $2.506 \times 10^{8}$ & $2.388\times 10^{8}$ & $1.94\times 10^8$ \\ \bottomrule
\end{tabular}
\end{table*}
\ra{1.0}

\begin{figure*}
\begin{subfigure}[b]{0.48\textwidth}
\centering
\includegraphics[height=0.4\linewidth]{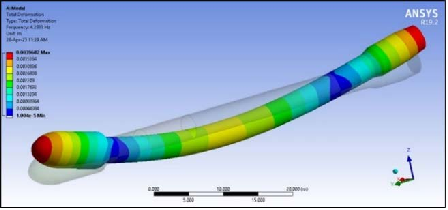}
\caption{First bending mode}
\label{fig:1stbending}
\end{subfigure}
\hfill
\begin{subfigure}[b]{0.48\textwidth}
\centering
\includegraphics[height=0.4\linewidth]{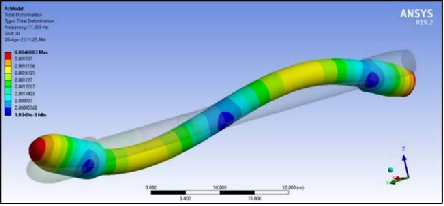}
\caption{Second bending mode}
\label{fig:2ndbending}
\end{subfigure}
\caption{First two flexible modes}
\label{fig:BendingShapes}
\end{figure*}

\begin{figure*}
    \centering
    \includegraphics[trim=50 0 90 0,clip, width=0.95\linewidth]{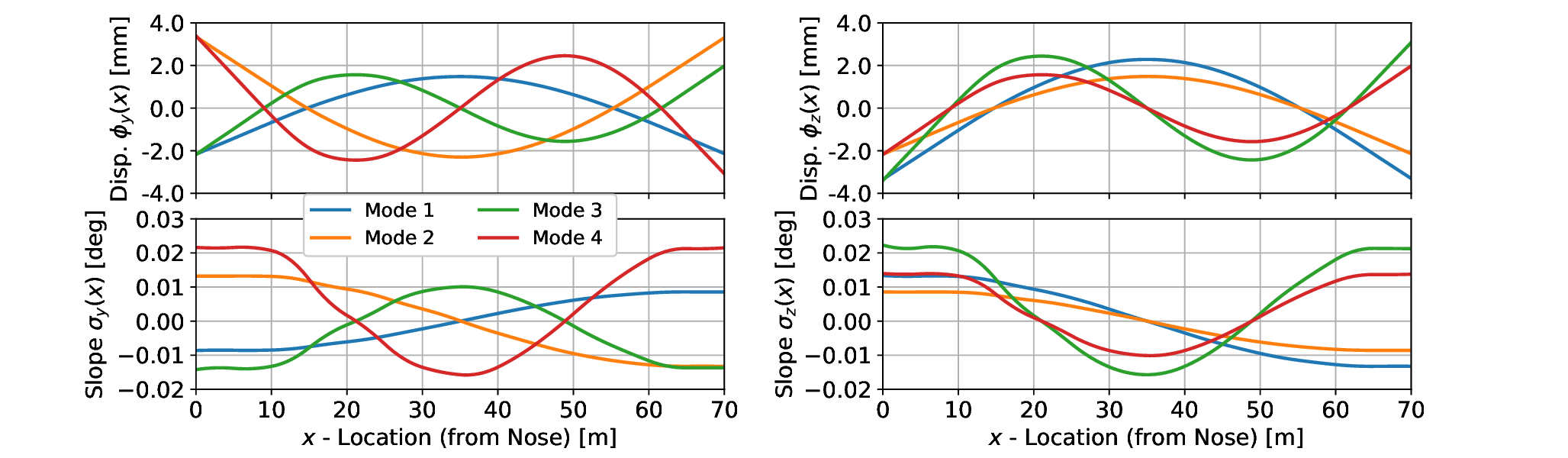}
    \caption{Variation of displacement and slope along the length of SLV}
    \label{fig:disp&slope}
\end{figure*}

Since the mass of space launch vehicles changes rapidly as the propellent burns during the flight, the value of inertia and the location of center of gravity also change at each point. Using CATIA the inertia matrix and the location of CG is determined for the CAD model. These parameters are calculated at the different fuel percentages and are tabulated in Table \ref{table:InertialData}. Interpolating these values gives us continuously changing inertial data corresponding to fuel percentage. This is achieved by decreasing the mass of fuel tank in the CAD model as shown in Fig. \ref{fig:ExplodedView}.

\subsection{Structural Analysis}\label{subsec:StrucAnalysis}
Before flexible body modeling, structural analysis needs to be done to determine flexible modes through modal analysis. Ansys Modal workbench is used to first generate a mesh for a simplified Falcon 9 model and then modal analysis is performed with both-ends-free boundary condition.

Figure \ref{fig:BendingShapes} shows an exaggerated view of the shape that the SLV takes under free vibrations when the first and second mode bending frequency are excited individually. 
Our focus lies on the effects that accrue due to these first two bending modes. The frequencies for the two modes are tabulated in Table \ref{tab:freq1&2}.

\ra{1.3}
\begin{table}
    \centering
    \caption{Frequency of 1\textsuperscript{st} and 2\textsuperscript{nd} Bending Mode}
    \label{tab:freq1&2}
    \begin{tabular}{cccc}\toprule
    \textbf{Mode}  & \textbf{Mode Type} & \textbf{Frequency (Hz)} & \textbf{Damping Ratio}\\
    \cmidrule(lr){1-4}
    7\textsuperscript{th} & 1\textsuperscript{st} Bending Mode & 4.293 & 0.0145\\
    8\textsuperscript{th} & 1\textsuperscript{st} Bending Mode & 4.293 & 0.0145\\
    9\textsuperscript{th} & 2\textsuperscript{nd} Bending Mode & 11.559 & 0.0147\\
    10\textsuperscript{th} & 2\textsuperscript{nd} Bending Mode & 11.559 & 0.0147\\ \bottomrule
    \end{tabular}
\end{table}
\ra{1.0}

To calculate the mode shapes, i.e. modal displacements and their slopes, at the above-mentioned frequencies, a python library \emph{pyAnsys} \cite{pyAnsys} is used, which provides tools to import and analyze Ansys output files. Damping ratio for each flexible mode is estimated using classical Rayleigh damping method and using historical values from literature \cite{Nesline1985}. Thus, the displacement and slope of each node along the length of the SLV are computed for first two modes, in both $y$ and $z$ axis, and are shown in Fig. \ref{fig:disp&slope}. It shows modal displacements ($\phi$) and slopes $(\sigma)$ in $y$ and $z$ directions, along the $x$ axis. Two locations along the $x$-axis are important, the location of nozzle at $70$ m from nose and the location of sensors at $15$ m from nose. These specific locations are denoted by subscripts $T$ and $G$, respectively, e.g. $\phi_{Y_T}$ represents  value of $\phi_Y$ at 70 m, and $\sigma_{Z_G}$ represents  value of $\sigma_Z$ at 15 m, etc. This assumed sensor location is where deflection is minimum for first mode excitement.

\subsection{Rigid Body Dynamics}
Considering the assumptions described earlier, and noting that time variations of mass and inertia matrix are significant, we can write the 6-DOF equation of motion of rigid dynamics as follows,
\begin{equation}\label{eq:RigidDyn}
\begin{split}
m (\dot{V} + \omega \times V) &= m g R^{\B_r}_{\I} e_3 + F_{aero} + F_T \\
\dot{J}\omega + J \dot{\omega} + \omega \times J \omega &= M_{aero} + \tau \\
\end{split}
\end{equation}
where $e_3 = [0,0,1]^\top$, $V = [u,v,w]^\top$ represents body velocity, $\omega = [p,q,r]^\top$ represents body angular velocity, $m$ and $J$ are SLV's mass and inertia matrix, respectively. $F_{aero}$ and $M_{aero}$ represents the aerodynamic forces and moments, respectively.  Moreover, $F_T$ and $\tau$ represents the forces and moments due to all engines and their gimbal deflections, respectively.

\begin{figure}
\begin{subfigure}[b]{0.41\linewidth}
\centering
\includegraphics[width=\linewidth]{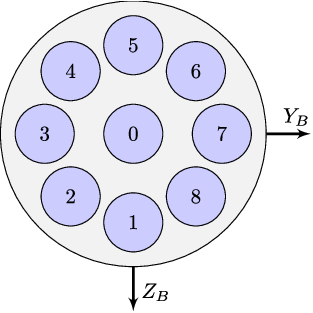}
\caption{Engines configuration}
\end{subfigure}
\hfill
\begin{subfigure}[b]{0.57\linewidth}
\centering
\includegraphics[width=\linewidth]{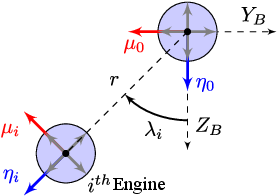}
\caption{Gimbal angles}
\end{subfigure}
\caption{Engines configuration and gimbal angles convention}
\label{fig:EngineGimbals}
\end{figure}

First stage of the Falcon 9 SLV is powered by nine \emph{Merlin} engines \cite{falcon9guide}, each of them is equipped with 2D gimballed nozzles. Consider the schematic shown in Fig. \ref{fig:EngineGimbals}, we can write total engine forces and moment as follows,
\begin{equation}\label{eq:ContForces}
\begin{split}
F_T &= \sum_{i=0}^{8}F_i(\delta_i) \\
\tau &= \sum_{i=0}^{8} \begin{bmatrix}- (L - x_{cg}) \\  -r \sin\lambda_i \\ r \cos\lambda_i \end{bmatrix} \times F_i(\delta_i)
\end{split}
\end{equation}
where, $\delta=[\delta_0^\top,\cdots,\delta_8^\top]^\top$, $\delta_i = [\mu_i,\eta_i]^\top$ represents the gimbal deflections of $i$th engine, for all $i\in[0,8]$, $L$ represents the length of SLV, and
\begin{equation}\label{eq:engineForce}
F_i(\delta_i) = R_{e_1} (\lambda_i) R_{e_3}(\mu_i) R_{e_2} (\eta_i) T_i e_1,
\end{equation}
where $T_i$ is the thrust of $i$th engine, and is assumed to be same for all engine, and it varies along the altitude as follows \cite{falcon9guide},
\begin{equation}\label{eq:thrust}
T_i = 914.11 - 0.68 P\;\;[\mathrm{kN}],\;\qquad \forall\;i\in[0,8]
\end{equation}
where is $P$ represents the atmospheric pressure in kPa at a given altitude.

\subsection{Flexible Modes Dynamics}
In reality, a flexible structure contains an infinite degrees of freedom, making an exact analysis nearly impossible. However, by restricting the system to a finite number of degrees of freedom, an approximate analysis can be accomplished. We employed a finite element method (FEM) based strategy to reduce the degrees of freedom. FEM model was developed in Ansys Modal workbench as discussed in section \ref{subsec:StrucAnalysis}, only first four modes (first two bending modes) are considered \cite{Nesline1985}, and the \emph{mass normalized} mode shapes are obtained. Restricting to these selected modes, we can write flexible dynamics as follows,
\begin{equation}\label{eq:FlexDyn}
\begin{bmatrix} \dot{\xi} \\ \ddot{\xi} \end{bmatrix} = \begin{bmatrix} 0 & I\\ -\Omega^2 & -2\zeta\Omega \end{bmatrix} \begin{bmatrix} \xi \\ \dot{\xi} \end{bmatrix}
+ \phi_T \sum\limits_{i=0}^8 F_i(\delta_i)
\end{equation}
where $\xi\in\R^{n_f}$ represents normalized deflection, $F_i(\delta_i)$ as in Eq. \eqref{eq:engineForce}, $\Omega = \mathrm{diag}(\Omega_1,\Omega_2,\cdots,\Omega_{n_f})$, $\zeta = \mathrm{diag}(\zeta_1,\zeta_2,\cdots,\zeta_{n_f})$. Where $\Omega_j$ and $\zeta_j$ represents the modal frequency and damping of $j$th flexible mode for all $j$ in $[1,n_f]$, here $n_f$ is the number of flexible modes considered. In this paper we have selected $n_f = 4$, and
\begin{equation}
\phi_T = \begin{bmatrix}
    0 & \phi_{Y_T} & \phi_{Z_T}
\end{bmatrix} \in \R^{n_f \times 3}
\end{equation}
where $\phi_{Y_T}$ and $\phi_{Z_T}$ represents the mode shapes as defined in section \ref{subsec:StrucAnalysis}.

\subsection{Interactions of Flexible and Rigid Modes}
To develop the complete nonlinear model of flexible SLV, we also need to model the effects of flexibility on rigid dynamics. We followed a similar approach as in \cite{Blakelock1991,barrows2020}, and considered the following effects.
\subsubsection{Flexibility Effects on Engine Gimbals}
Structural flexibility affects gimbal deflections. Forces and moments due to all nine engines gets modified due the bending effects. For that total engine forces and moments in Eq. \eqref{eq:RigidDyn} $F_T$ and $\tau$ will get replaced by $\hat{F}_T$ and $\hat{\tau}$, respectively, which are expressed as follows,
\begin{equation}
\begin{split}
\hat{F}_T &= \sum_{i=0}^8  \hat{F}_i(\delta_i)\\
\hat{\tau} &= \sum_{i=0}^{8} \left(\begin{bmatrix}- (L - x_{cg}) \\  -r \sin\lambda_i \\ r \cos\lambda_i \end{bmatrix}  + \phi_T^\top \xi  \right) \times \hat{F}_i(\delta_i)
\end{split}
\end{equation}
where,
\begin{equation}
\hat{F}_i(\delta_i) = R_{e_3}(\sigma_{Y_T}^\top \xi)R_{e_2}(\sigma_{Z_T}^\top \xi)F_i(\delta_i)
\end{equation}
where, $F_i(\delta_i)$ is the $i$th engine force vector without bending as described in Eq. \eqref{eq:engineForce}, and $\sigma_{Y_T}$ and $\sigma_{Z_T}$ are mode slopes at engine gimbal location as defined in section \ref{subsec:StrucAnalysis}.

\subsubsection{Flexibility Effects on Sensor Measurements}
Another key effect of flexible modes is their contribution in sensor measurements, which if not properly taken care of, can get fed back into the system through control and may results in an unstable positive feedback loop. Let $\B_f$ be the local frame at sensor location, and aligned with the body frame $\B_r$ in the absence of bending. Then in case of bending we can write,
\begin{equation}
R_\I^{\B_f} = R_{\B_r}^{\B_f} R_\I^{\B_r}
\end{equation}
With small angle assumption, i.e. bending effects are small in magnitude, and using Rodrigues' formula we can write $R_{\B_r}^{\B_f}$ as follows
\begin{equation}
R_{\B_r}^{\B_f} \approx \begin{bmatrix}
    1 & -\sigma_{Y_G}^\top \xi & \sigma_{Z_G}^\top \xi\\
    \sigma_{Y_G}^\top \xi & 1 & 0\\
    -\sigma_{Z_G}^\top \xi & 0 & 1
\end{bmatrix}
\end{equation}
Similarly, we can write body rates measured by the gyroscopes $\omega_m$ as follows,
\begin{equation}
\omega_m = \sigma_G^\top \dot{\xi} +  R_{\B_r}^{\B_f} \omega \approx \sigma_G^\top \dot{\xi} + \omega
\end{equation}
where,
\begin{equation}
\sigma_G = \begin{bmatrix} 0 & \sigma_{Z_G} & \sigma_{Y_G} \end{bmatrix} \in \mathbb{R}^{n_f \times 3}
\end{equation}
where $\sigma_{Z_G}$ and $\sigma_{Y_G}$ are the modal slopes at sensor location. It is worth noting that since only bending modes are considered, roll angle and roll rate aren't affected by flexibility.

\subsection{Control Allocation}
Assuming same thrust for all engines ($T_i=T$), and small gimbal deflections ($\delta$) we can linearize Eq. \eqref{eq:ContForces} as,
\begin{equation}
\tau \approxeq T \Lambda G \delta \triangleq T \Lambda \begin{bmatrix} \delta_A \\ \delta_E \\ \delta_R \end{bmatrix}
\end{equation}
where,
\[
\Lambda = \begin{bmatrix}
            8r & 0 & 0 \\
            0 & -9(L-x_{cg}) & 0 \\
            0 & 0 & -9(L-x_{cg})
          \end{bmatrix}
\]
and $G=[G_0,G_1\cdots,G_8]$, where $G_i = \frac{1}{T}\Lambda^{-1} \left(\frac{\partial \tau}{\partial \delta_i}\right)$, for all $i\in[0,8]$.
So, using pseudo-inverse we can write the control allocation as,
\begin{equation}
\delta = G^\dag \begin{bmatrix} \delta_A \\ \delta_E \\ \delta_R \end{bmatrix}
\end{equation}
where $G^\dag = G^\top\left(GG^\top\right)^{-1}$.

\subsection{Linearized Dynamics}\label{sec:LinDyn}
In the process of control system design for the flexible space launch vehicle, it is necessary to obtain a set of linear model. In this regard, we linearized the complete flexible dynamics discussed in previous subsections, and assumed the decoupling between different channels. The resulting short period approximation of longitudinal model is same as presented in \cite{Blakelock1991}, as described below.
\begin{equation}\label{eq:PitchLinearModel}
\begin{split}
\dot{\alpha} &= \frac{Z_{\alpha}}{V}\alpha + q + \frac{Z_{\delta_E}}{V} \delta_E + \sum_{r=1}^{n_f}\frac{Z_{\delta}\sigma_T^{(r)}}{V} \xi_r\\
\dot{q} &= M_{\alpha}\alpha + M_{\delta_E} \delta_E + \sum_{r=1}^{n_f}\left(M_{\delta_E}\sigma_T^{(r)} + \frac{m Z_\delta\phi_T^{(r)}}{I_y}\right) \xi_r \\
\dot{\theta} &= q \\
\ddot{\xi}_r &= -\omega_r^2 \xi_r  -2\zeta\omega_r \dot{\xi}_r + mZ_{\delta_E}\sigma_T^{(r)} \delta_E, \quad \forall\;r\in[1,n_f]
\end{split}\qquad\raisetag{3.5\baselineskip}
\end{equation}

Moreover, since the effects of flexibility on roll dynamics are negligible, therefore, for roll channel, linear models are same as that of rigid dynamics and are shown below.
\begin{equation}\label{eq:RollLinearModel}
\begin{split}
\dot{\phi} &= p \\
\dot{p} &= L_p p + L_{\delta_A} \delta_A
\end{split}
\end{equation}

%% file: Sections/Control_Design.tex
\begin{figure}
    \centering
    \includegraphics[width=\linewidth]{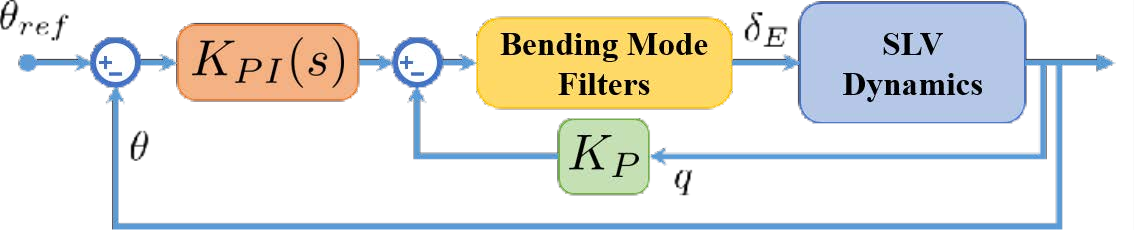}
    \caption{Control architecture}
    \label{fig:Control Architecture}
\end{figure}

In this section linear control design is presented. The linearized equations presented in section \ref{sec:LinDyn}, which account for the impact of flexibility effects, are used. These equations capture the short period dynamics of the flexible space launch vehicle while considering the influence of its inherent flexibility. Along a trajectory similar to that used in \cite{Youtube2022} a set of linear models are developed for different angle of attack conditions. Control architecture for pitch and yaw channels is shown in Fig. \ref{fig:Control Architecture}. Similar architecture but without any bending mode filters is used for roll channel. As pitch and yaw channels are symmetric, so only pitch controller is presented. Moreover, since roll dynamics and its control design is trivial so it is skipped.

The first step in a control design process is to select a single linear model, from of the set of linear models. For that purpose, the design point for the controller is selected where the system is most unstable. The rigid alone, and rigid and flexible combined transfer functions are shown in Eqs. \eqref{eq:TFrigid} and \eqref{eq:TFflex}, respectively. By incorporating four bending modes at two distinct frequencies, the transfer function for the pitch angle undergoes a significant transformation, as shown in Eq. \eqref{eq:TFflex}. The resulting transfer function now includes eight additional poles, further complicating the dynamics of the system. The inclusion of these bending modes in the transfer function allows for a more accurate representation of the flexible effects present in the space launch vehicle. By capturing the dynamics associated with bending modes at different frequencies, we gain a more comprehensive understanding of the system's behavior and can design a controller that effectively addresses these additional complexities.
\begin{equation}\label{eq:TFrigid}
\frac{\theta(s)}{\delta_E(s)} = \frac{-0.017725 (s+0.02853)}{s (s-0.4067) (s+0.4557)}
\end{equation}

\begin{equation}\label{eq:TFflex}
\frac{\theta(s)}{\delta_E(s)} = \frac{\begin{aligned}
-0.0142&(s-193.4) (s+184.2) (s+11.44) \\
&\;\;\;\;(s-11.58) (s+0.02848)\end{aligned}} {\begin{aligned}
s&(s+0.4557) (s^2 + 0.7826s + 727.6)\\
&\;\;(s-0.4067)(s^2 + 2.14s + 5275)\end{aligned}}
\end{equation}

\begin{figure}
    \centering
    \includegraphics[width=0.8\linewidth]{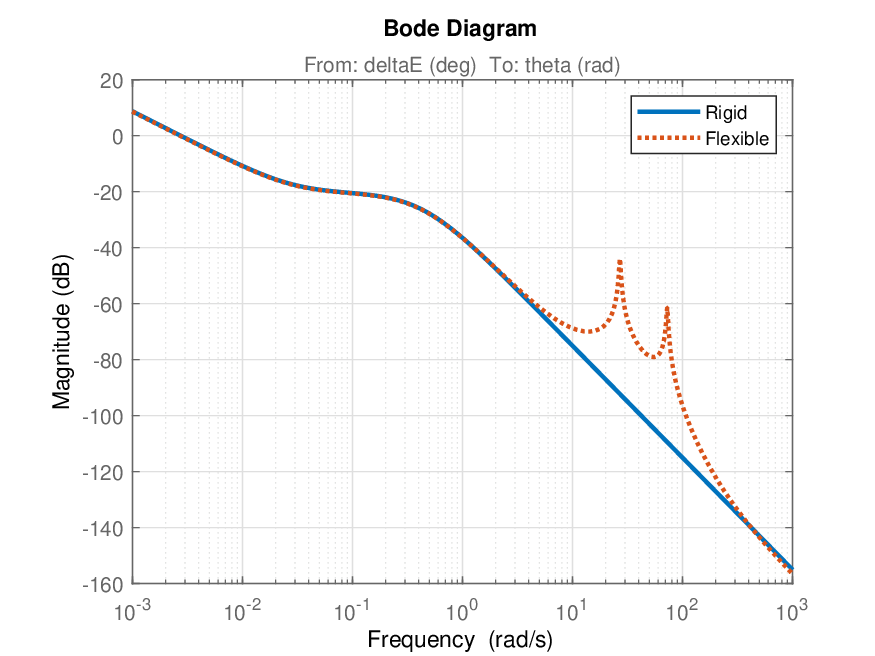}
    \caption{Comparison of rigid and flexible dynamics}
    \label{fig:CompRigidFlexBode}
\end{figure}

The comparison of the bode plots of transfer functions in Eqs. \eqref{eq:TFrigid} and \eqref{eq:TFflex} are shown in Fig. \ref{fig:CompRigidFlexBode}. Observing the Bode plots, distinct peaks can be observed in the red line for the flexible body. These peaks correspond to the presence of two bending modes that are not adequately attenuated. The presence of these peaks signifies potential instability in the system, as they introduce significant resonance and amplification at specific frequencies associated with the bending modes.

To address the destabilizing peaks in the frequency response of the system, the incorporation of filters into the controller is required. In the introduction section, various filters have been mentioned. Among the available filter types, the elliptic filter stands out as an ideal choice for our system due to its ability to provide a sharp cut-off and effectively attenuate specific frequencies associated with the bending modes \cite{Samar2008}. Moreover, elliptic filter has wide stopband allowing for effective rejection of frequencies outside the desired passband. On the other hand, notch filters provide an efficient solution in terms of phase lag if modal frequencies are accurately known. Therefore, in this work we considered both elliptic and notch filters and compared their performance. For notch filter we used a double notch whose transfer function is shown below,
\begin{equation}
G_{N}(s) = \frac{(s^2 + 0.27s + 727.4) (s^2 + 0.73s + 5275)}{(s^2 + 37.76s + 727.4) (s^2 + 101.7s + 5275)}
\end{equation}

The elliptic filter parameters are selected as follows:
\begin{itemize}
    \item First order (n) is 3.
    \item Passband Frequency (Wp) is 10 rad/s.
    \item Passband ripple (Rp) is 1 dB.
    \item Stopband sttenuation (Rs) is 40 dB.
\end{itemize}
These values gives the following elliptic filter using MATLAB \texttt{ellip} command,
\begin{equation}
G_{E}(s) = \frac{0.69201 (s^2 + 760.8)}{(s+5.237) (s^2 + 4.545s + 100.5)}
\end{equation}

Controller gains and compensator are tuned, and following values were selected,
\begin{itemize}
  \item $K_P = -114.5916$
  \item $K_{PI}(s) = -214.2862\left(1 + \frac{0.1}{s}\right)$
\end{itemize}

\begin{figure}
    \centering
    \includegraphics[width=0.8\linewidth]{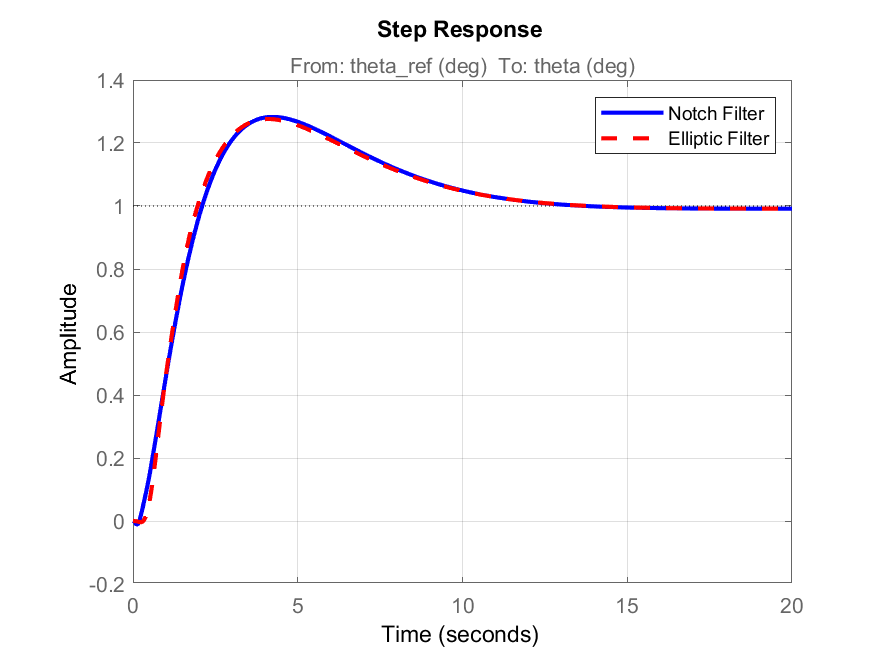}
    \caption{Step response comparison: Notch and Elliptic filters}
    \label{fig:StepResponseComp}
\end{figure}

\begin{figure}
    \centering
    \includegraphics[width=0.8\linewidth]{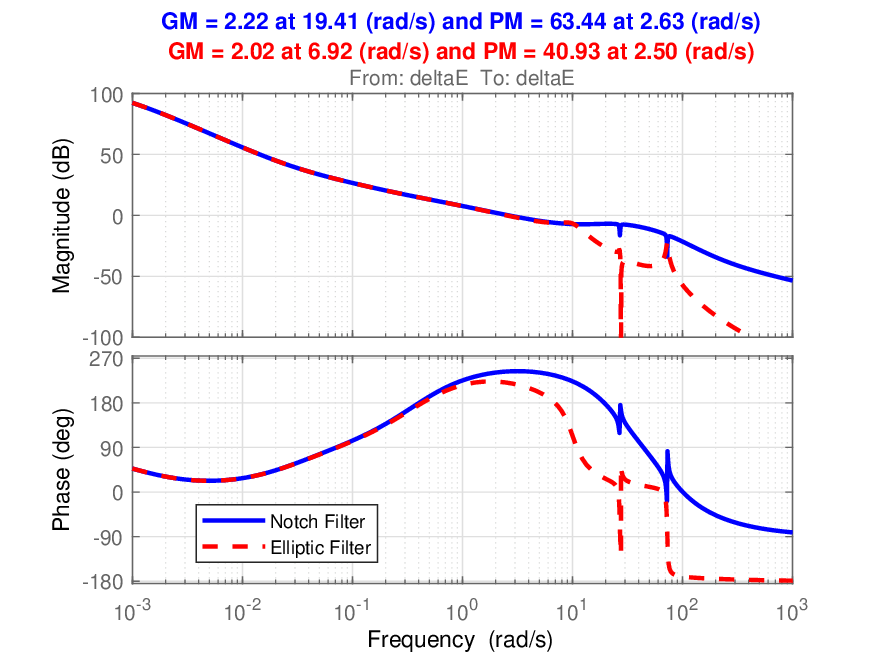}
    \caption{Frequency response comparison: Notch and Elliptic filters}
    \label{fig:BodeResponseComp}
\end{figure}

This controller, along with both filters separately, was analyzed on flexible models of the space launch vehicle. Fig. \ref{fig:StepResponseComp} shows comparison of step responses with both filters and it can be seen that they are almost similar. However, form loop shape bode plot comparison shown in Fig. \ref{fig:BodeResponseComp}, we can see that the notch filter provides better gain and phase margins as compared to the elliptic filter.

%% file: Sections/Nonlinear_Simulation.tex
\begin{figure}
    \centering
    \includegraphics[trim=140 200 140 200, clip, width=0.8\linewidth]{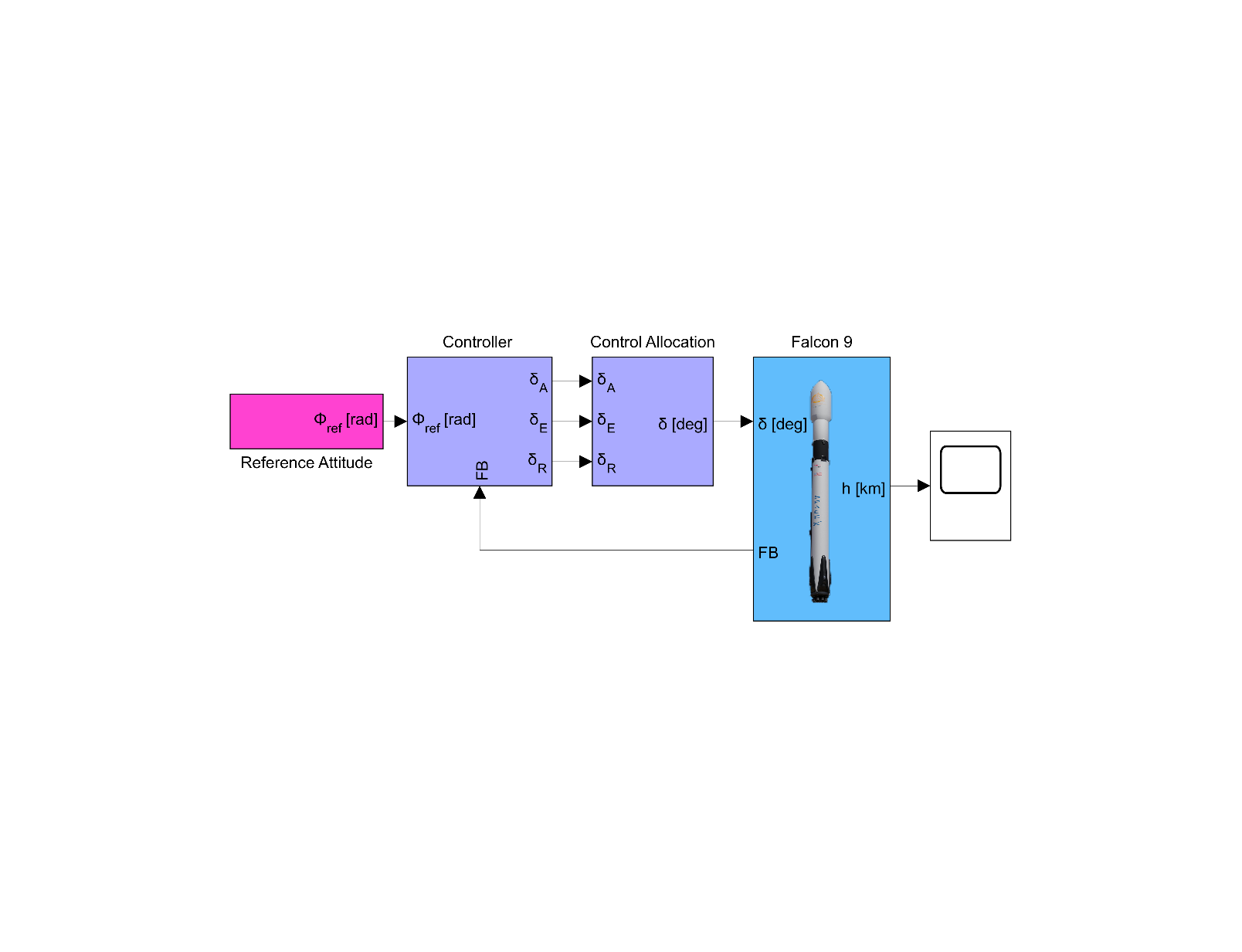}
    \caption{Nonlinear simulation}
    \label{fig:Complete Simulink Diagram}
\end{figure}
\begin{figure*}
    \centering
    \includegraphics[trim=0 120 0 130, clip, width=0.85\linewidth]{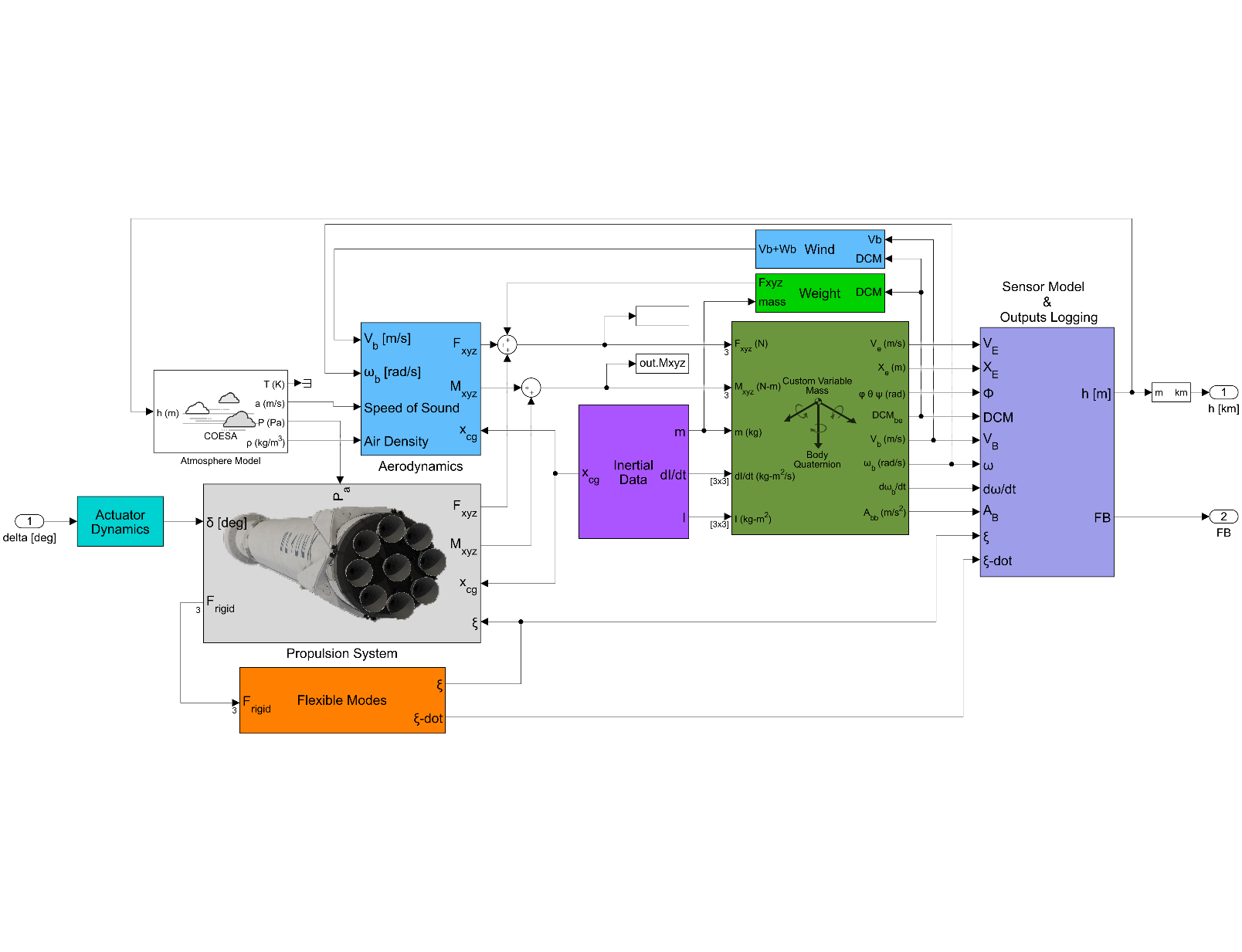}
    \caption{Falcon 9 dynamics block}
    \label{fig:Falcon 9 Dynamics Block}
\end{figure*}
The designed controller and filter are implemented in nonlinear simulation developed in Simulink as shown in Fig. \ref{fig:Complete Simulink Diagram}. The dynamics block of the Falcon 9 SLV is shown in Fig. \ref{fig:Falcon 9 Dynamics Block}. A reference trajectory for pitch and yaw angle is to be followed by Falcon 9. For pitch angle trajectory it is assumed that the SLV remains completely vertical for the first 10 seconds and then pitch angle decreases linearly from $90^\circ$ to $40^\circ$ for the rest of the trajectory. Similarly for yaw angle it is assumed that Falcon 9 is on a trajectory to the \emph{International Space Station}, thus the launch azimuth angle required from \emph{Cape Canaveral} is $135^\circ$. Only trajectory, similar to that in \cite{Youtube2022}, till first stage separation is considered, that is about 165 seconds.

\begin{figure}
    \centering
    \includegraphics[width=0.85\linewidth]{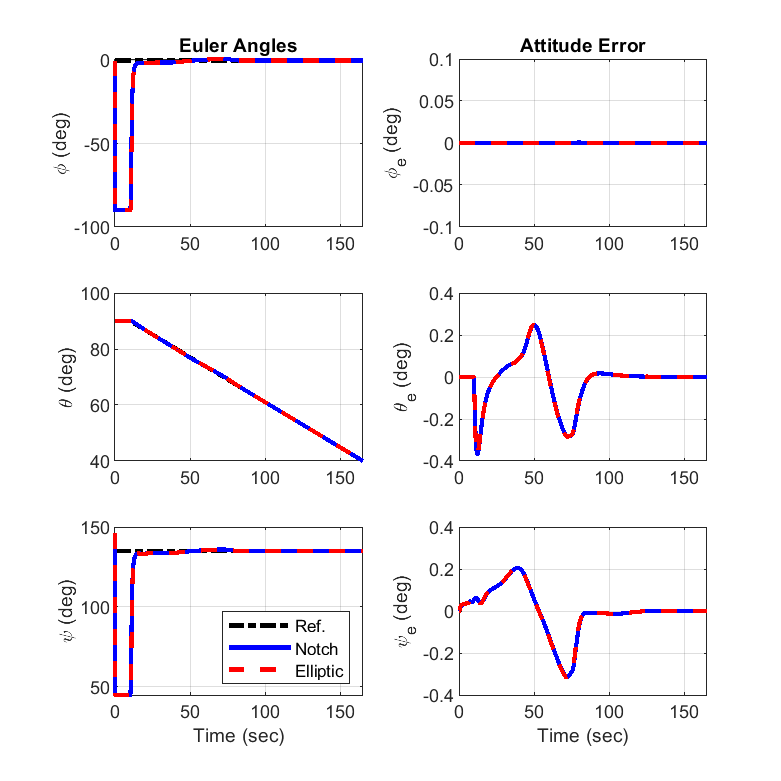}
    \caption{Nominal performance: Euler angles and attitude errors}
    \label{fig:NomResults_EulerAngle}
\end{figure}
\begin{figure}
    \centering
    \includegraphics[width=0.7\linewidth]{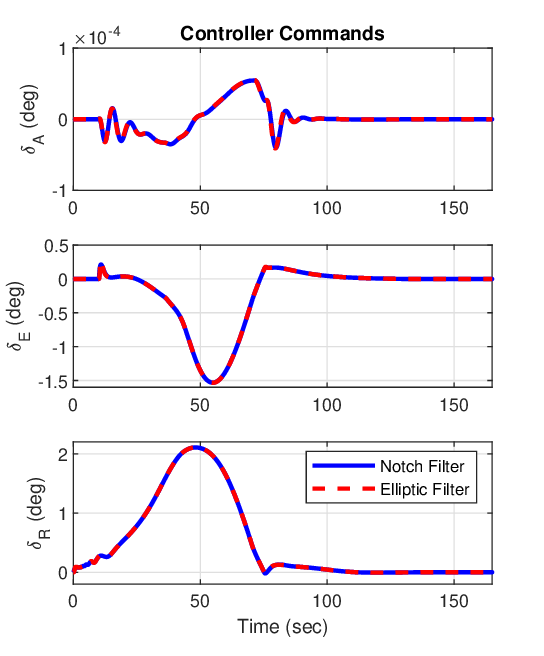}
    \caption{Nominal performance: Controller commands}
    \label{fig:NomResults_ControlCmds}
\end{figure}

The nonlinear simulation results for both Notch and Elliptic filters, are shown in Figs. \ref{fig:NomResults_EulerAngle} and \ref{fig:NomResults_ControlCmds}. In these simulations, nominal values of all parameters are considered, and wind of 10 knots is applied along north and east direction. Fig. \ref{fig:NomResults_EulerAngle} depicts the attitude angle and it can be seen that the SLV follows the reference trajectory for pitch and yaw angle quite accurately, and the attitude errors remains within a fraction of a degree over the complete trajectory. Similarly, as shown Fig. \ref{fig:NomResults_ControlCmds} controller commands are also small. Moreover, both filters have same the performance in the nominal scenario.

To compare the robustness of controller with each filter towards the uncertainty in modal parameters, i.e. mode frequencies and mode shapes, Monte-Carlo type simulations were performed. Despite their similar nominal performance, with Elliptic filter closed loop remained stable upto $\pm 34\%$ variation in modal parameters, while with Notch filter it was stable only upto $\pm 3\%$ variations. This observation, in contrast to fact that controller with notch filters has more gain and phase margins, is consistent with the results in \cite{Samar2008}. Which further emphasize that the elliptic filters are preferable for gain stabilization, while simple notch filters should be used only when modal parameters are precisely known.